\documentclass[useAMS,usenatbib]{mn2e}
\usepackage{multirow}
\usepackage{epsfig}
\usepackage{subfigure}

\makeatletter
\let\jnl@style=\rm
\def\ref@jnl#1{{\jnl@style#1}}

\def\aj{\ref@jnl{AJ}}                   
\def\araa{\ref@jnl{ARA\&A}}             
\def\apj{\ref@jnl{ApJ}}                 
\def\apjl{\ref@jnl{ApJ}}                
\def\apjs{\ref@jnl{ApJS}}               
\def\ao{\ref@jnl{Appl.~Opt.}}           
\def\apss{\ref@jnl{Ap\&SS}}             
\def\aap{\ref@jnl{A\&A}}                
\def\aapr{\ref@jnl{A\&A~Rev.}}          
\def\aaps{\ref@jnl{A\&AS}}              
\def\azh{\ref@jnl{AZh}}                 
\def\baas{\ref@jnl{BAAS}}               
\def\jrasc{\ref@jnl{JRASC}}             
\def\memras{\ref@jnl{MmRAS}}            
\def\mnras{\ref@jnl{MNRAS}}             
\def\pra{\ref@jnl{Phys.~Rev.~A}}        
\def\prb{\ref@jnl{Phys.~Rev.~B}}        
\def\prc{\ref@jnl{Phys.~Rev.~C}}        
\def\prd{\ref@jnl{Phys.~Rev.~D}}        
\def\pre{\ref@jnl{Phys.~Rev.~E}}        
\def\prl{\ref@jnl{Phys.~Rev.~Lett.}}    
\def\pasp{\ref@jnl{PASP}}               
\def\pasj{\ref@jnl{PASJ}}               
\def\qjras{\ref@jnl{QJRAS}}             
\def\skytel{\ref@jnl{S\&T}}             
\def\solphys{\ref@jnl{Sol.~Phys.}}      
\def\sovast{\ref@jnl{Soviet~Ast.}}      
\def\ssr{\ref@jnl{Space~Sci.~Rev.}}     
\def\zap{\ref@jnl{ZAp}}                 
\def\nat{\ref@jnl{Nature}}              
\def\iaucirc{\ref@jnl{IAU~Circ.}}       
\def\aplett{\ref@jnl{Astrophys.~Lett.}} 
\def\apspr{\ref@jnl{Astrophys.~Space~Phys.~Res.}}
\def\bain{\ref@jnl{Bull.~Astron.~Inst.~Netherlands}}
\def\fcp{\ref@jnl{Fund.~Cosmic~Phys.}}  
\def\gca{\ref@jnl{Geochim.~Cosmochim.~Acta}}   
\def\grl{\ref@jnl{Geophys.~Res.~Lett.}} 
\def\jcp{\ref@jnl{J.~Chem.~Phys.}}      
\def\jgr{\ref@jnl{J.~Geophys.~Res.}}    
\def\jqsrt{\ref@jnl{J.~Quant.~Spec.~Radiat.~Transf.}}
\def\memsai{\ref@jnl{Mem.~Soc.~Astron.~Italiana}}
\def\nphysa{\ref@jnl{Nucl.~Phys.~A}}   
\def\physrep{\ref@jnl{Phys.~Rep.}}   
\def\physscr{\ref@jnl{Phys.~Scr}}   
\def\planss{\ref@jnl{Planet.~Space~Sci.}}   
\def\procspie{\ref@jnl{Proc.~SPIE}}   

\makeatother

\title[The XMM-\textit{Newton} view of Mrk~3 and IXO~30]{The XMM-\textit{Newton} view of Mrk~3 and IXO~30}
\author[Stefano Bianchi, Giovanni Miniutti, Andrew C. Fabian, Kazushi Iwasawa]{Stefano Bianchi$^{1}$ $^2$ $^3$ \thanks{E-mail: Stefano.Bianchi@sciops.esa.int (SB)}, Giovanni Miniutti$^3$, Andrew C. Fabian$^3$, Kazushi Iwasawa$^3$\\
$^{1}$XMM-Newton Science Operations Center, European Space Astronomy Center, ESA, Apartado 50727, E-28080 Madrid, Spain\\
$^2$Dipartimento di Fisica, Universit\`a degli Studi Roma Tre, Via della Vasca Navale 84, I-00146, Roma, Italy\\
$^3$Institute of Astronomy, Madingley Road, Cambridge, CB3 0HA\\
}

\begin{document}

\pagerange{\pageref{firstpage}--\pageref{lastpage}} \pubyear{2004}

\maketitle

\label{firstpage}

\begin{abstract}

We present the analysis of the XMM-\textit{Newton} EPIC pn spectrum of the Seyfert 2 galaxy, Mrk~3. We confirm that the source is dominated by a pure Compton reflection component and an iron K$\alpha$ line, both produced as reflection from a Compton-thick torus, likely responsible also for the large column density ($1.36^{+0.03}_{-0.04}\times10^{24}$ cm$^{-2}$) which is pierced by the primary powerlaw only at high energies.
A low inclination angle and an iron underabundance of a factor $\simeq0.82$, suggested by the amount of reflection and the depth of the iron edge, are consistent with the iron K$\alpha$ line EW with respect to the Compton reflection component, being $610^{+30}_{-50}$ eV. Moreover, the iron line width, $\sigma=32^{+13}_{-14}$ eV, if interpreted in terms of Doppler broadening due to the Keplerian rotation of the torus, puts an estimate to the inner radius of the latter, $r=0.6^{+1.3}_{-0.3}\,\sin^2{i}$ pc. Finally, two different photoionised reflectors are needed to take into account a large number of soft X-ray emission lines from N, O, Ne, Mg, Si, Fe L and the {Fe\,\textsc{xxv}} emission line at $6.71^{+0.03}_{-0.02}$ keV. RGS spectra show that the soft X-ray spectrum is dominated by
emission lines, while the underlying continuum is best fitted by an unabsorbed powerlaw with the same photon index of the primary continuum, produced as reflection by a photoionised material with a column density of a few $10^{22}$ cm$^{-2}$.
We also present the first X-ray spectrum of \textit{ROSAT} source IXO~30, which shows a huge iron line at $6.5^{+0.3}_{-0.2}$ keV and is well represented either by an absorbed powerlaw with $\Gamma\simeq1.8$ or bremsstrahlung emission at a temperature of $7.5^{+2.1}_{-1.6}$ keV. Its spectral properties point to a likely identification in terms of a weak Galactic Cataclysmic Variable, but the lack of any optical counterpart precludes excluding other possibilities, like an ULX at the distance of Mrk~3.

\end{abstract}


\section{Introduction}

Mrk~3 (z=0.0135) is a prototypical Seyfert 2 galaxy, the optical broad lines of which are seen only in polarised light \citep{mg90,tran95}. As soon as it was observed in X-rays by \textit{GINGA}, it was clear that the source was heavily absorbed and presented a strong iron K$\alpha$ line \citep{awaki90,awaki91}. Moreover, a soft excess was detected by \textit{Einstein} \citep{kruper90} and \textit{ROSAT} \citep{tum93}. \citet{iwa94} suggested that it could originate in an extended region, likely dominated by scattering of the intrinsic continuum, on the basis of the \textit{ASCA} observation, which also indicated a significant flux variability above 4 keV with respect to the previous data.

The availability of broad band observations, due to \textit{RXTE} \citep{georg99} and, in particular, \textit{BeppoSAX} \citep{cappi99}, allowed these authors to disentangle three different components: an heavily absorbed powerlaw, an unabsorbed Compton reflection component associated to the iron line and a constant soft X-ray emission. The latter component was later found to be spatially extended along the [{O\,\textsc{iii}}] cone and likely produced in a warm photoionised material, as shown by the high spatial and energy resolution observation by \textit{Chandra} \citep{sako00b}.

In this paper, we analyse the XMM-\textit{Newton} EPIC pn and the RGS spectra of Mrk~3, comparing the results with the long X-ray history of the source, with the aim to estimate some properties of the material around its nucleus.

\section{Data redution}

\subsection{XMM-\textit{Newton}}

Mrk~3 was observed by XMM-\textit{Newton} between 2000 October 19 and 20, with all the EPIC CCD cameras, the pn \citep{struder01} and the two MOS \citep{turner01}, operating in Full Frame and Medium filter. This paper deals only with pn data, the count rate of which is well below the maximum for 1 per cent pileup (see Table 3 of the XMM-Newton Users' Handbook). \textsc{SAS} 6.0.0 was used to reduce data, with a total net exposure time of 52 ks, after screening for intervals of flaring particle background, applying the procedure to maximise the signal-to-noise ratio introduced by \citet{pico04}. An extraction radius of 40 arcsec was used for the single and double pattern spectra. After a consistency check by fitting the two spectra separately, we decided to extract a single spectrum, with all patterns from 0 to 4.
A weak source is apparent in the pn image just outside our extraction region. However, its flux (measured from the \textit{Chandra} spectrum: see below) is completely negligible with respect to Mrk~3.
RGS1 and RGS2 spectra of Mrk~3 were also extracted, adopting standard procedures, with total exposure times of 60 and 58 ks, respectively.

Finally, another source lies at $\simeq1.6$ arcmin from the nucleus, corresponding to \textit{ROSAT} source IXO~30 \citep{cp02}. We extracted EPIC pn, MOS1 and MOS2 spectra for this object, adopting extraction radii of 19 arcsec, for total exposure times of 47, 49 and 52 ks, respectively, again after applying the \citet{pico04} procedure. In the following fits, the two MOS spectra were furthermore added into a single spectrum. Moreover, in order to study the flux variability of this source, we analysed all the available EPIC pn observations of the Mrk~3 field (listed in Table \ref{ixo30_obs}), with the only exception of observation \textsc{0009220901}, where IXO~30 lies partly in the chip gap.

\subsection{\textit{Chandra}}

\textit{Chandra} observed Mrk~3 with the ACIS-S HETG between 2000 March 18 and 19, for $\simeq100$ ks. The data were already analysed by \citet{sako00b}. However, in order to better compare them to pn data, we re-extracted first order HEG and MEG spectra with \textsc{Ciao} 3.1 and \textsc{Caldb} 2.27, following standard procedures.

The 0th order spectrum of a source ($\alpha_{2000}=06^h15^m29^s.7$, $\delta_{2000}=+71\degr 01\arcmin 40\arcsec .9$) lying at $\simeq45$ arcsec from the nucleus was extracted, to verify if it could contaminate the pn spectrum (see above). An extraction region with a radius of 3 arcsec was used. The resulting spectrum has a 2-10 keV flux of $\simeq2.5\times10^{-14}$ erg cm$^{-2}$ s$^{-1}$, thus being 200 times weaker than Mrk~3.

Finally, the 0th order spectrum of IXO~30 ($\alpha_{2000}=06^h15^m15^s$, $\delta_{2000}=+71\degr 02\arcmin 05\arcsec$) was also extracted, adopting an extraction region with a radius of 3 arcsec.\\

All spectra were analysed with \textsc{Xspec} 11.3.1. In the following, errors correspond to the 90\% confidence level for one interesting parameter ($\Delta \chi^2 =2.71$), where not otherwise stated. The cosmological parameters used throughout this paper are $H_0=70$ km s$^{-1}$ Mpc$^{-1}$, $\Lambda_0=0.73$ and $q_0=0$.

\section{Data analysis}
\label{analysis}

\subsection{Temporal behaviour}

\begin{figure*}
\begin{center}
\subfigure[] 
 {
    \label{mrk3_lc}
     \epsfig{file=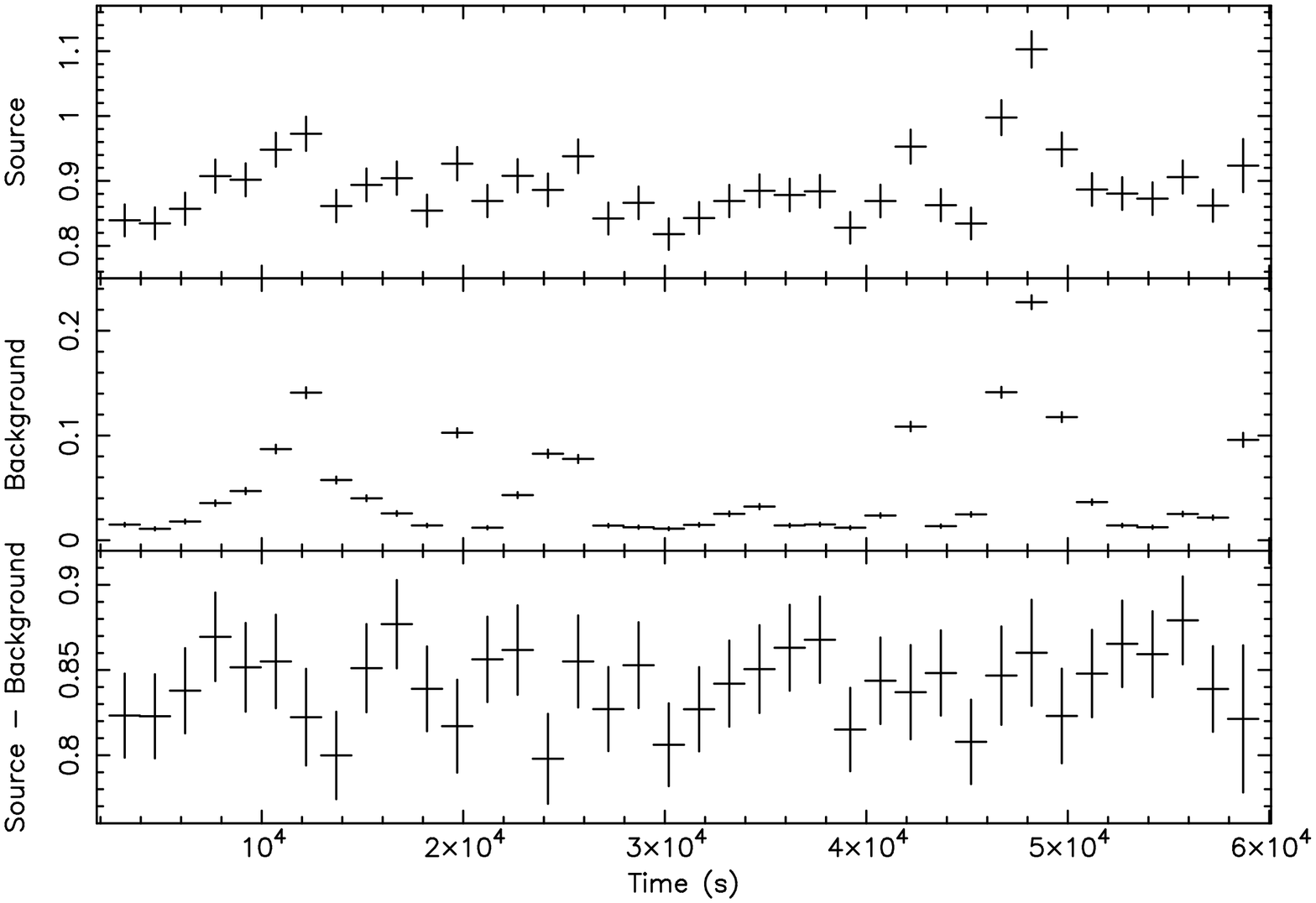, width=5.5cm, angle=-90}
}
\hspace{0.5cm}
\subfigure[] 
{
    \label{hardratio}
     \epsfig{file=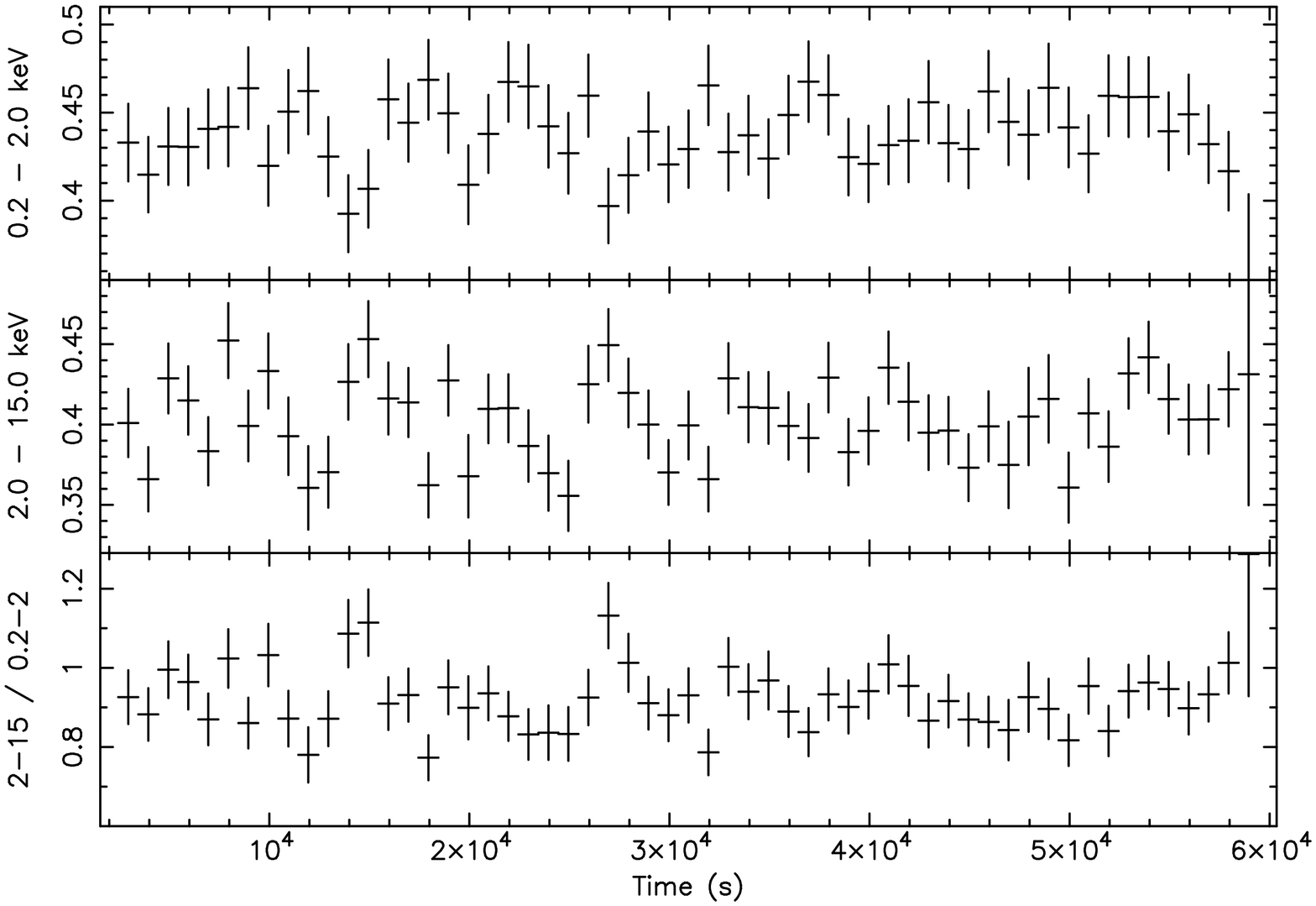, width=5.5cm, angle=-90}
}

\end{center}

\caption{\label{lightcurve}(a) \textit{Upper}: Lightcurve of the adopted extraction region for Mrk~3. \textit{Middle}: Lightcurve for the background, with count rates rescaled to the area of the source extraction region. \textit{Lower}: Background-subtracted lightcurve of Mrk~3. (b) \textit{Upper}: 0.2 - 2.0 keV background subtracted lightcurve for Mrk~3. \textit{Middle}: 2.0 - 15.0 keV background subtracted lightcurve for Mrk~3. \textit{Lower}: The ratio of the latter to the former.}

\end{figure*}

Figure \ref{mrk3_lc} shows the EPIC pn lightcurve of Mrk~3, along with the lightcurve of the background. We can conclude that there is no evidence for significant variability during the observation. Moreover, soft and hard X-rays lightcurves are plotted in Figure \ref{hardratio}, along with their ratio. Again, taking into account sistematic errors due to the subtraction of the background, no variability is apparent in the observation. As a consequence, we use the whole available exposure time in the following spectral analysis.

\subsection{Spectrum}

\begin{table}
\caption{\label{continuum}Best fit parameters for the continuum (see text for details.)}
\begin{center}
\begin{tabular}{cc}
\hline
$N_H$ (cm$^{-2}$) & $1.36^{+0.03}_{-0.04}\times10^{24}$\\
&\\
$\Gamma$ & $1.77\pm0.01$\\
&\\
$A_{Fe}$ & $0.82^{+0.10}_{-0.08}$\\
&\\
$\cos{i}$ & 0.45$^{a}$\\
&\\
$F_{\mathrm{0.5-2\,keV}}$ (erg cm$^{-2}$ s$^{-1}$) & $6.3\times10^{-13}$\\
&\\
$F_{\mathrm{2-10\,keV}}$ (erg cm$^{-2}$ s$^{-1}$) & $5.9\times10^{-12}$\\
&\\
$L_{\mathrm{0.1-150\,keV}}$ (erg s$^{-1}$) & $1.1\times10^{44}$ $^b$\\
&\\
$\chi^2$/d.o.f. & 249/208\\
\hline
\end{tabular}
\end{center}
$^a$ \textsc{Xspec} default value - $^b$Absorption-corrected extrapolated luminosity.
\end{table}

We fitted the 0.5-13 keV spectrum with the model adopted by \citet{cappi99} for the broadband BeppoSAX data, which includes a strongly absorbed powerlaw and the components arising from reprocessing of this primary continuum from a Compton-thick material, that is a pure Compton reflection with the same photon index \citep[model \textsc{pexrav}: ][]{mz95} and an iron K$\alpha$ line. Moreover, a steeper, unabsorbed, powerlaw is used to model the soft excess. A column density of $8.46\times10^{20}$ cm$^{-2}$ was included in the model, to take into account the Galactic absorption \citep{dl90}.

\begin{figure*}

\begin{center}
\subfigure[] 
 {
    \label{pnspectrum}
    \epsfig{file=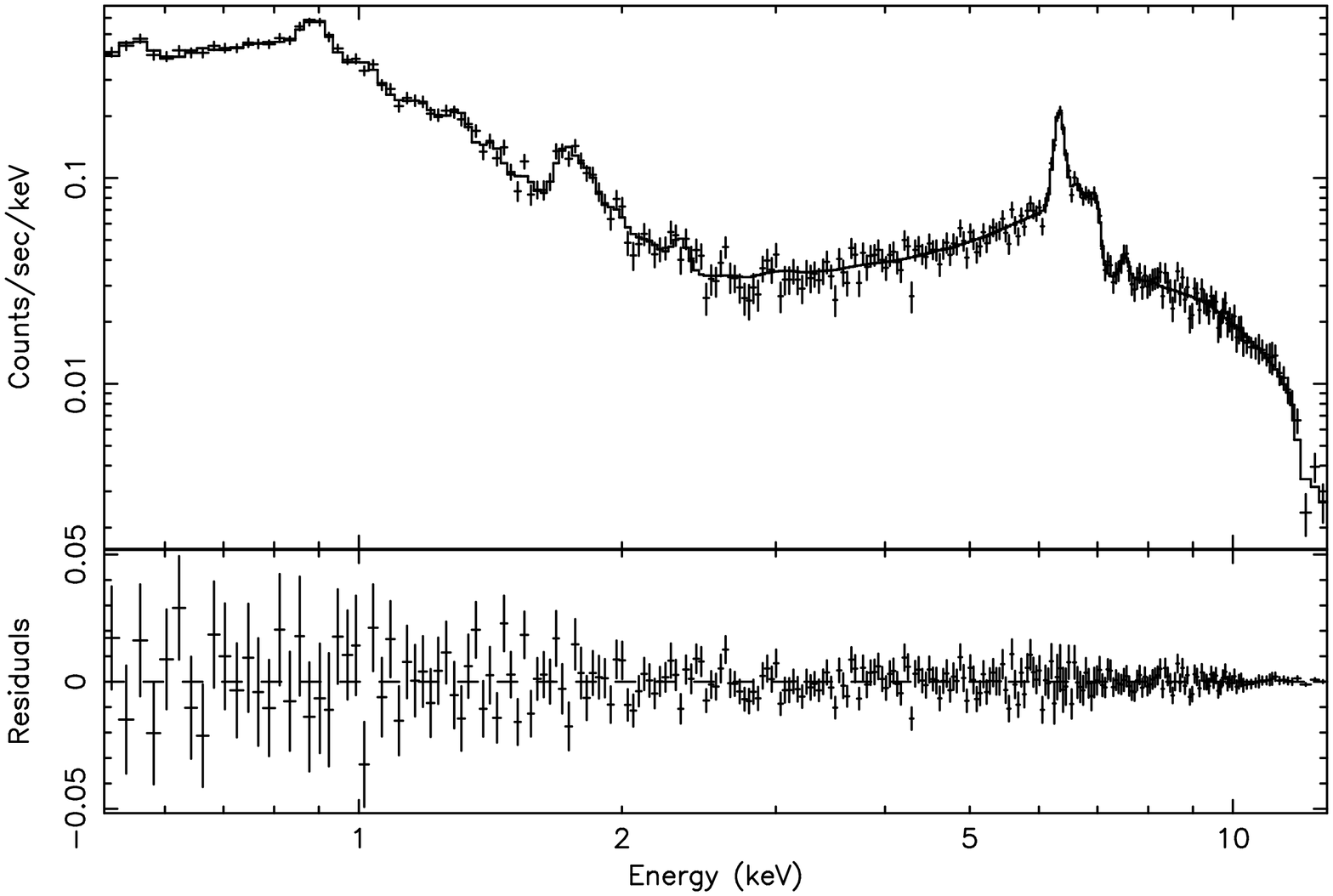, width=5.2cm, angle=-90}
}
\hspace{0.5cm}
\subfigure[] 
{
    \label{emodel}
    \epsfig{file=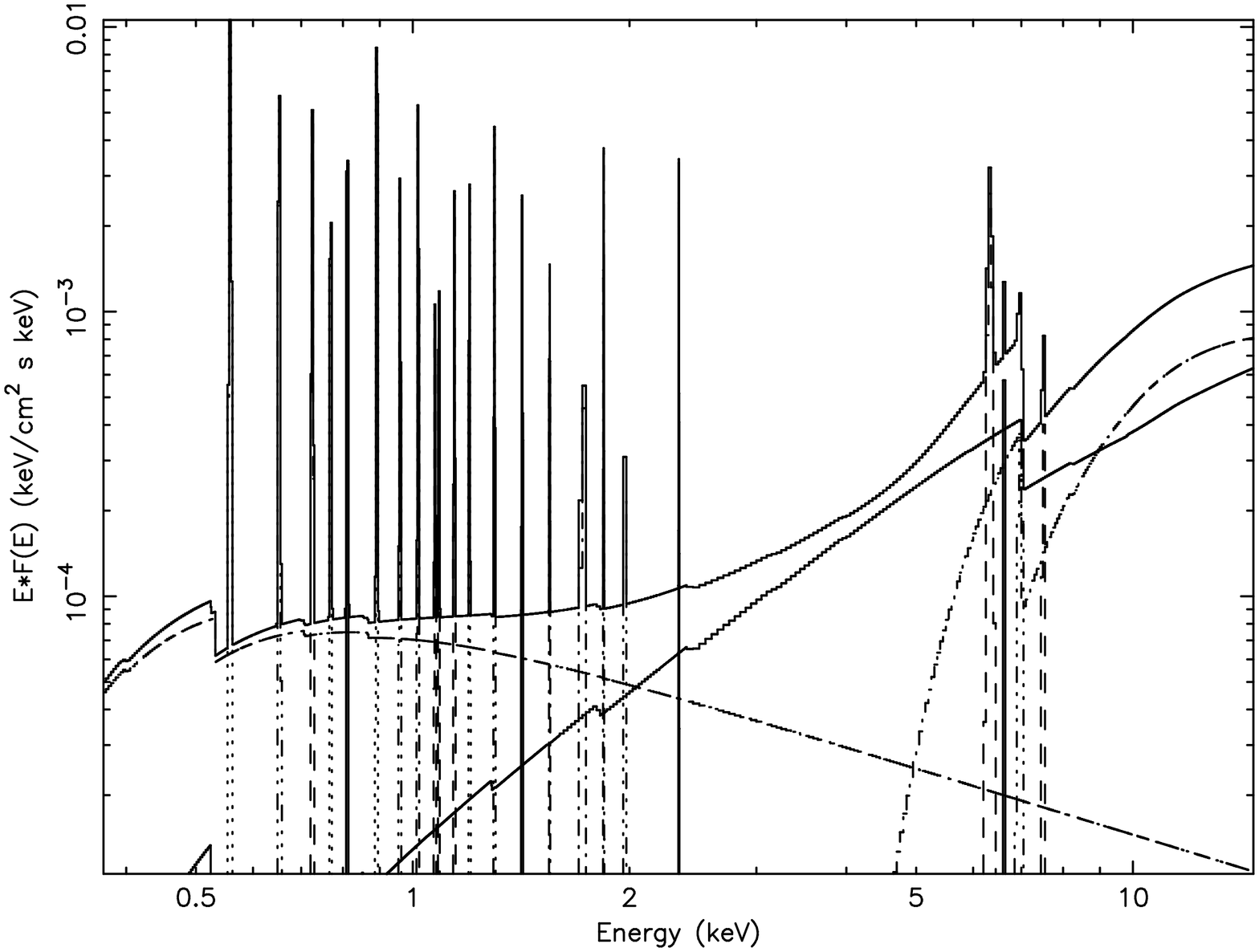, width=5.2cm, angle=-90}
}

\end{center}
\caption{\textit{(a)} Best fit model and residuals for the EPIC pn spectrum. \textit{(b)} The adopted model plotted as $E\cdot F(E)$.}

\end{figure*}

The resulting fit is unacceptable ($\chi^2=1074/251$ d.o.f.), mainly because of large residuals in the soft X-ray spectrum, likely due to emission lines. Moreover, the index of the softer powerlaw is quite steep ($\simeq3$), so that it is not easy to assess its physical origin. Therefore, we examined the RGS spectra, before improving the pn fit. We found that the soft X-ray spectrum is dominated by emission lines, in particular {O\,\textsc{vii}} and {O\,\textsc{viii}} K$\alpha$ lines between 0.5-0.7 keV. Including these lines greatly improves the soft X-ray fit of the lower resolution pn spectrum. We identified each single line with known transitions: when the identification was likely affected by a blend of lines, we tried to include each line in the model, but only when these features improved the reduced $\chi^2$. A powerlaw component is still required by the data, but now its photon index can be kept fixed to the one found for the primary continuum. The final fit is good ($\chi^2=249/208$ d.o.f.): the continuum parameters are shown in Table \ref{continuum}, while a detailed discussion on the emission line spectrum is deferred to Sect. \ref{discussion}. Fig. \ref{pnspectrum} and \ref{emodel} show the spectrum and the adopted best fit model.

The powerlaw index of the primary continuum and the local absorbing column density are well constrained, the two values being $1.77\pm0.01$ and $1.36^{+0.03}_{-0.04}\times10^{24}$ cm$^{-2}$. An interesting result from the fit is an hint for iron underabundance, the best fit value being $0.82^{+0.10}_{-0.08}$ \citep[with respect to the solar abundances measured by][]{ag89}. However, this result must be taken with caution, because this parameter is actually calculated on the basis of the iron edge depth of the Compton reflection continuum, which depends tightly also on the primary powerlaw index and the inclination angle.

\section{Discussion}
\label{discussion}

\subsection{\label{iron}The torus}

The Compton reflection component and the strong iron K$\alpha$ line are clear signatures of reflection from a Compton-thick, fairly neutral, material. It is then natural to assume that the absorber and the reflector are one and the same material, the torus, as usually found in other Compton-thick Seyfert galaxies. It is interesting to note that we also detected a line consistent with emission from Si less ionised than {Si\,\textsc{viii}} (see Table \ref{toruslines}), thus likely being another product of reflection from the torus. This feature was also found in the \textit{Chandra} spectrum \citep{sako00b}.

\subsubsection{\label{CR}The Compton reflection}

The 2-10 keV flux of the Compton reflection component \textit{only} is $3.4\times10^{-12}$ erg cm$^{-2}$ s$^{-1}$, completely consistent with the $3.3\times10^{-12}$ erg cm$^{-2}$ s$^{-1}$ measured by the \textit{Chandra} observation, performed seven months before. To check for variability in past X-ray observations, we can compare the 2-10 keV \textit{total} flux reported in Table \ref{continuum}, which by far is dominated by the reflection component (see Fig. \ref{emodel}). The 1997 BeppoSAX observation measured a flux of $6.5\times10^{-12}$ erg cm$^{-2}$ s$^{-1}$ \citep{cappi99}, slightly larger than the EPIC pn one, the difference probably lying in the brighter intrinsic continuum of the source at that time (see below). Indeed, \citet{cappi99} performed a detailed comparison between their dataset and the \textit{GINGA} and \textit{ASCA} observations, concluding that there was evidence of a variable component above the iron line energy and a constant one below.

This is also supported by some indications that the intrinsic, unabsorbed luminosity (0.1-150 keV) of Mrk~3 changed between the XMM-\textit{Newton} and \textit{Chandra} observations and with respect to the BeppoSAX one, the values being $1.1\times10^{44}$, $0.8\times10^{44}$ and $1.3\times10^{44}$ erg s$^{-1}$, respectively\footnote{The latter two values are those reported by \citet{sako00b} and \citet{cappi99}, corrected to have the same choice of cosmological parameters adopted in this paper.}. However, it should be stressed that this luminosity is driven by the normalization of the strongly absorbed powerlaw, so the measures with XMM-\textit{Newton} and, in particular, \textit{Chandra} are much less reliable than the one obtained with the broadband spectrum of BeppoSAX.

Finally, an estimate of the angle $i$ can be made on the basis of the amount of Compton reflection with respect to the incident nuclear continuum which is quite large \citep[$R\simeq1$:][]{cappi99}, thus requiring an inclination not very high to avoid self-obscuration of the torus.

\subsubsection{The iron K$\alpha$ line}

\begin{table}
\caption{\label{toruslines}List of the emission lines included in the best fit model, originating from a fairly neutral material, likely the torus.}
\begin{center}
\begin{tabular}{cccc}
\textbf{Energy (keV)} & \textbf{Flux$^a$} & \textbf{Id.} & $\mathbf{E_T}$ $^b$ \textbf{(keV)}\\
\hline
& & & \\
$1.744^{+0.018}_{-0.002}$ & $0.7\pm0.1$ &{Si\,\textsc{i-viii}} K$\alpha$ & 1.740-1.769\\
& & & \\
$6.415\pm0.006$ & $3.8\pm0.2$ & {Fe\,\textsc{xiii-xvi}} K$\alpha$ & 6.411-6.424\\
& & & \\
$7.06^*$ & $0.5\pm0.2$ &{Fe\,\textsc{i-xvi}} K$\beta$ & 7.058\\
& & & \\
\hline
\end{tabular}
\end{center}
$^a$ $10^{-5}$ ph cm$^{-2}$ s$^{-1}$

$^b$ Theoretical energies for the transitions \citep[data from NIST: ][ and references intherein]{nist}.

$^*$ Fixed
\end{table}

A prominent iron K$\alpha$ line is present around 6.4 keV. However, its EW with respect to the Compton reflection component, being $610^{+30}_{-50}$ eV, falls short of the expected one, which should be larger than 1 keV \citep[see e.g.][]{mbf96}. Two factors can contribute in reconciling the observed value with the theoretical one. One is a small inclination angle, since the iron line EW increases with this parameter. The other is iron underabundance: the dependence of EW with this parameter is almost linear for $A_{Fe}<1$ \citep{mbf96}. Both these requirements are satisfied by the Compton reflection component properties, which arises in the same material: the value of $R$ is consistent with low inclination angles (see Sect. \ref{CR}) and the iron edge depth suggests an iron underabundance of a factor $\simeq0.82$ (see Sect. \ref{analysis}).

The iron K$\alpha$ width is actually resolved in the EPIC pn spectrum, with $\sigma=32^{+13}_{-14}$ eV. Even if this measure should be taken with care because it is well below the instrument resolution, it is interesting to note that it is fully compatible with the value found by \textit{Chandra}, which is $\sigma=28^{+11}_{-7}$ eV. The observed iron line width can be in principle explained by several means.

One effect that should be in principle taken into account is that the neutral iron line is indeed composed of a doublet,  K$\alpha_1$ at 6.404 keV and K$\alpha_2$ at 6.391 keV, with a flux ratio of 2:1 \citep{bea67}. The difference between the two lines, being lower than the spectral resolution of present X-ray detectors, is generally ignored and a weighted mean of 6.400 keV is commonly used. \citet{yaq01} already noted that a fit with two Gaussian lines does not vary significantly the measure of the total line width in \textit{Chandra} gratings data. We performed the same test with our high statistics EPIC pn spectrum, fitting the neutral iron line with a K$\alpha_1$ and a K$\alpha_2$, whose energies and relative intensities were freezed to the atomic values, while their width was constrained to vary together. The resulting, common width is $\sigma=35^{+14}_{-16}$ eV, in full agreement with the value found with a single Gaussian line\footnote{Note that in order to achieve a fit statistically equivalent to the one with a single Gaussian line (whose best fit energy was significantly higher than that of neutral iron), we froze the energy separation between the two transitions (13 eV, appropriate for neutral iron), but allowed the K$\alpha_1$ energy to vary. The resulting best fit value for the K$\alpha_1$ energy is around 6.418 keV, i.e. well above the one expected for {Fe\,\textsc{ix}} \citep[see e.g.][ for the theoretical energies of the iron doublet for some Fe ions]{palm03}.}.

On the other hand, since the line is produced by reprocessing from a Compton-thick material, a Compton Shoulder (CS) is expected on theoretical grounds \citep{matt02}, as a result of Compton scattering of the line photons on the same material where they originate. Indeed, the iron line profiles of Circinus and NGC~1068 are successfully modelled with a narrow core and a CS both in XMM-\textit{Newton} and \textit{Chandra} spectra \citep{bianchi02,mbm03,matt04,ogle03}, with a better statistical significance with respect to a single Gaussian line with a resolved width. Therefore, another possibility is that the observed FWHM is the result of a bad modelling of the overall profile, which does not include the CS. To parametrise this component, we added a further Gaussian line, with centroid energy at 6.3 keV and $\sigma=40$ eV \citep{matt02}. The CS is not required on statistical grounds ($\Delta\chi^2\simeq0$ for 1 d.o.f. less), but the CS flux, even if low, is consistent with production from a Compton-thick material, being $10^{+9}_{-6}$\% that of the line core, and the iron line width is now unresolved (the upper limit being 50 eV).
We then tested if this model applies well also to the \textit{Chandra} HETG data. However, this is not the case. An unresolved line core plus a CS gives a significantly worse fit with respect to a single, resolved Gaussian ($\Delta\chi^2=13$ for the same d.o.f.). On the other hand, if the line core width is left free to vary (with a best fit value of $\sigma=29^{+10}_{-8}$), we get an upper limit to the CS flux of 13\%, still compatible with the expectations, but not explaining the FWHM of the line.

A more likely possibility is that we are not observing a line from purely neutral iron. If this is the case, the iron K$\alpha$ line is expected to be actually the blend of a number of emission features from different ionic species, each composed by a doublet, whose transitions are separated by energies of tens eV. Indeed, the best fit value of the centroid energy for the line in the EPIC pn spectrum suggests that iron is ionised in the range {Fe\,\textsc{xiii-xvi}} \citep{house69}. However, the $K\beta/K\alpha$ flux ratio is $0.12^{+0.06}_{-0.05}$. This value does not depend only on the ratio between the fluorescence yields of the two transitions (1:8), but also on the different total flux of the continuum and photoabsorption cross sections at the energies of the two lines. To account for all these effects, \citet{mbm03} used the \cite{basko78} formul\ae\ and obtained for neutral iron a ratio of 0.155-0.160, depending on the inclination angle. Therefore, the measured value in Mrk~3, albeit smaller than the expected value, is fully consistent with a production from neutral iron. Moreover, the K$\beta$ fluorescent yield decreases with the ionisation stage, until it becomes null for {Fe\,\textsc{xvii}}, when the M shell is completely void and the transition is impossible. Therefore, taking into account the above-mentioned value for neutral iron and scaling it with the fluorescent yields calculated by \citet{km93}, we can estimate that the observed value for the $K\beta/K\alpha$ flux ratio is appropriate for {Fe\,\textsc{x-xi}}, but is inconsistent with iron more ionised than {Fe\,\textsc{xii-xiii}}, just marginally in agreement with the centroid energy of the line as observed by the pn.

Finally, if we assume that the torus rotates around the Black Hole (BH) with a Keplerian velocity, it is easy to show that the expected FWHM for a line produced in its inner walls is approximately $2\,v_\mathrm{k}\,\sin{i} \simeq 1\,300 \left(M_8/r_{\mathrm{pc}} \right) ^{\frac{1}{2}}\,\sin{i}$ km s$^{-1}$, where the mass is in units of 10$^8$ M$_\odot$, the inner radius in parsec and $i$ is the angle between the torus axis and the line of sight. The BH mass of Mrk~3 is estimated by means of stellar velocity
dispersion to be $4.5\times10^8$ M$_\odot$ \citep{wu02}.
This means that the expected FWHM for the iron line in this source is $2\,770\,r_{\mathrm{pc}}^{-1/2}\,\sin{i}$ km s$^{-1}$.
Therefore, the line width we measure with the EPIC pn, corresponding to FWHM=$3\,500^{+1\,400}_{-1\,500}$ km s$^{-1}$ if due to Doppler broadening, puts an estimate on the inner radius of the torus: $r=0.6^{+1.3}_{-0.3}\,\sin^2{i}$ pc. A choice of $30\degr$ or $60\degr$ leads to central values of 0.3 and 0.5 pc respectively for the inner radius.

\begin{figure}

\epsfig{file=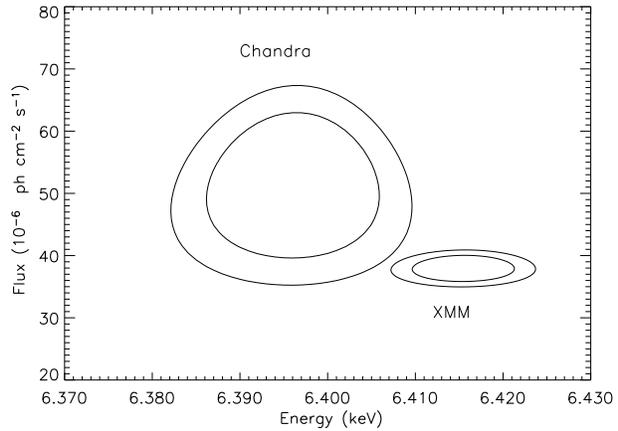, width=8cm}

\caption{\label{energyflux}Energy vs. flux contour plot for the `neutral' iron K$\alpha$ line.}

\end{figure}

Moreover, there is a marginal indication that the line flux and energy actually changed between the \textit{Chandra} and the XMM-\textit{Newton} observation. In Fig. \ref{energyflux} the contour plots for the iron line energy vs. flux are shown for both spectra, showing that the two curves are not consistent with each other at the 90\% confidence level (for two interesting parameters). Even if the reason for this discrepancy is more likely to be found in a less than perfect calibration of the two instruments \citep[as it may be the case in other high statistics EPIC pn spectra of AGN, see e.g.][]{mbm03,matt04}, it may be a good exercise to understand if such a variation is indeed possible, by estimating the expected photoionisation time for the neutral iron present in the torus.

The photoionisation time can be expressed in a simplified way, as done by \citet{rey95}. The formula they estimated for {O\,\textsc{vii}} can be used also in our case, rescaling for the photoinisation cross section and the threshold energy appropriate for neutral iron \citep[see][]{vy95}, so that:

\begin{equation}
t_{ph}\simeq1\,200\,R_{16}^2\,L_{43}^{-1}\,s
\end{equation}

where $R=10^{16}\,R_{16}$ cm is the distance of the material from a source with ionising luminosity ($E>13.6$ eV) of $L=10^{43}\,L_{43}$ erg s$^{-1}$. Adopting an approximate ionising luminosity of $10^{44}$ erg s$^{-1}$ (see Table \ref{continuum}), we find that, for a distance of the torus of 1 pc, the resulting $t_{ph}$ is of the order of $10^7$ s, i.e. approximately 4 months. This means that a variation of the photoionisation structure of the inner wall of the torus is perfectly reasonable in a time span of 7 months, which separates the two observations.

\subsection{\label{ionized}The ionised reflectors}

\subsubsection{The soft X-ray spectrum}

\begin{table*}
\caption{\label{lines}List of the emission lines originating from an ionised material included in the best fit model, along with their fluxes and their likely identifications (see text for details).}
\begin{center}
\begin{tabular}{ccccccc}
\multicolumn{2}{c}{\textbf{EPIC pn}} & \multicolumn{2}{c}{\textbf{RGS}} & & \\
\textbf{Energy (keV)} & \textbf{Flux$^a$} & \textbf{Energy (keV)} & \textbf{Flux$^a$} & \textbf{Id.} & $\mathbf{E_T}$ $^b$ \textbf{(keV)}\\
\hline
\multirow{3}*{--} & \multirow{3}*{--} & \multirow{3}*{$0.43^{+0.03}_{-0.02}$} & \multirow{3}*{$2.7^{+1.9}_{-1.2}$} & \multirow{3}*{{N\,\textsc{vi}} K$\alpha$} & 0.420 (f)\\
& & & & & 0.426 (i)\\
& & & & & 0.431 (r)\\
& & & & &\\
-- & -- & $0.50^{+0.02}_{-0.03}$ & $1.4^{+1.2}_{-0.4}$ & {N\,\textsc{vii}} K$\alpha$ & 0.500 \\
& & & & &\\
\multirow{3}*{$0.565^{+0.003}_{-0.004}$} & \multirow{3}*{$14.4^{+1.0}_{-0.8}$} & $0.562\pm0.001$ & $6.3^{+1.3}_{-2.2}$ & {O\,\textsc{vii}} K$\alpha$ & 0.561 (f)\\
& & \multirow{2}*{$0.57\pm0.01$} & \multirow{2}*{$2.5^{+1.4}_{-1.3}$} & \multirow{2}*{{O\,\textsc{vii}} K$\alpha$} & 0.569 (i)\\
& & & & & 0.574 (r)\\
& & & & &\\
\multirow{2}*{$0.661^{+0.005}_{-0.004}$} & \multirow{2}*{$6.2^{+0.5}_{-0.4}$} & $0.655\pm0.005$ & $3.2^{+0.6}_{-0.8}$ & {O\,\textsc{viii}} K$\alpha$ & 0.654\\
& & $0.666^*$ & $0.5^{+0.7}_{-0.4}$ & {O\,\textsc{vii}} K$\beta$ & 0.666\\
& & & & &\\
$0.735^{+0.005}_{-0.006}$ & $3.7^{+0.5}_{-0.3}$ & $0.741^{+0.005}_{-0.008}$ & $2.8^{+0.7}_{-1.1}$ & {O\,\textsc{vii}} RRC & $>0.739$\\
& & & & &\\
$0.780^{+0.009}_{-0.011}$ & $1.9^{+0.4}_{-0.3}$ & $0.775^*$ & $0.8^{+0.4}_{-0.7}$ & {O\,\textsc{viii}} K$\beta$ & 0.775 \\
& & & & &\\
$0.822\pm0.006$ & $3.2\pm0.3$ & $0.827\pm0.002$ & $1.1^{+0.7}_{-0.3}$ &  {Fe\,\textsc{xvii}} L & 0.826\\
& & & & &\\
\multirow{4}*{$0.904^{+0.003}_{-0.002}$} & \multirow{4}*{$6.4\pm0.3$} & $0.88^{+0.01}_{-0.02}$ &  $1.6\pm0.4$ &
{O\,\textsc{viii}} RRC & $>0.871$\\
& &  \multirow{2}*{$0.91^{+0.01}_{-0.03}$} & \multirow{2}*{$1.8^{+0.4}_{-0.8}$} & \multirow{2}*{{Ne\,\textsc{ix}} K$\alpha$} & 0.905 (f)\\
& & & & & 0.915 (i)\\
& & $0.924\pm0.002$ & $1.8^{+0.4}_{-1.0}$ & {Ne\,\textsc{ix}} K$\alpha$ & 0.922 (r)\\
& & & & &\\
\multirow{2}*{$0.973\pm0.009$} & \multirow{2}*{$1.4^{+0.3}_{-0.2}$} & \multirow{2}*{$1.00^{+0.01}_{-0.02}$} & \multirow{2}*{$1.4^{+0.5}_{-0.8}$} & {Fe\,\textsc{xx}} L & 0.965\\
& & & & {Fe\,\textsc{xxi}} L & 1.009\\
& & & & &\\
$1.031\pm0.005$ & $2.5\pm0.2$ & $1.02^{+0.05}_{-0.01}$ & $1.9^{+0.4}_{-0.9}$ & {Ne\,\textsc{x}} K$\alpha$ & 1.022\\
& & & & &\\
$1.09^{+0.02}_{-0.03}$ & $0.4\pm0.2$ & $1.08^{+0.01}_{-0.03}$ & $1.0^{+0.8}_{-0.6}$ & {Fe\,\textsc{xxii}} L & 1.053\\
& & & & &\\
$1.10^{+0.02}_{-0.01}$ & $0.6^{+0.2}_{-0.1}$ & $1.128^*$ & $<0.5$ & {Fe\,\textsc{xxiii}} L & 1.128\\
& & & & &\\
$1.16\pm0.01$ & $0.9^{+0.2}_{-0.1}$ & $1.170^*$ & $0.8^{+0.8}_{-0.4}$ & {Fe\,\textsc{xxiv}} L & 1.170\\
& & & & &\\
$1.215^{+0.010}_{-0.007}$ & $1.0\pm0.1$ & $1.24^{+0.02}_{-0.04}$ & $0.8^{+0.6}_{-0.3}$ & {Ne\,\textsc{x}} K$\beta$ & 1.211\\
& & & & &\\
\multirow{3}*{$1.316^{+0.007}_{-0.006}$} & \multirow{3}*{$1.3\pm0.1$} & \multirow{2}*{$1.34^{+0.01}_{-0.04}$} & \multirow{2}*{$0.8^{+0.7}_{-0.3}$} &\multirow{2}*{{Mg\,\textsc{xi}} K$\alpha$} & 1.331 (f)\\
& & & & & 1.344 (i)\\
& & $1.36^{+0.03}_{-0.04}$ & $0.7^{+0.7}_{-0.3}$ & {Mg\,\textsc{xi}} K$\alpha$ & 1.352 (r)\\
& & & & &\\
$1.44\pm0.01$ & $0.7\pm0.1$ & $1.47^{+0.01}_{-0.02}$ & $0.8^{+0.8}_{-0.4}$ & {Mg\,\textsc{xii}} K$\alpha$ & 1.472 \\
& & & & &\\
$1.57^{+0.02}_{-0.03}$ & $0.3\pm0.1$ & -- & -- & {Mg\,\textsc{xi}} K$\beta$ & 1.578 \\
& & & & &\\
\multirow{3}*{$1.866^{+0.012}_{-0.003}$} & \multirow{3}*{$0.6\pm0.1$} & \multirow{3}*{--} & \multirow{3}*{--} &\multirow{3}*{{Si\,\textsc{xiii}} K$\alpha$} & 1.840 (f) \\
& & & & &1.855 (i)\\
& & & & &1.865 (r)\\
& & & & &\\
$1.99^{+0.03}_{-0.02}$ & $0.2\pm0.1$ & --& --&{Si\,\textsc{xiv}} K$\alpha$ & 2.005\\
& & & & &\\
$2.37^{+0.04}_{-0.03}$ & $0.4\pm0.1$ & --& --&{Si\,\textsc{xiv}} K$\beta$ & 2.376\\
& & & & &\\
\hline\\
\end{tabular}
\end{center}
$^a$ $10^{-5}$ ph cm$^{-2}$ s$^{-1}$\\
$^b$ Theoretical energies for the transitions. For He-like ions, forbidden, intercombination and resonant lines are separately listed, while the photoionisation threshold is listed in the case of RRC \citep[data from NIST: ][ and references intherein]{nist}\\
$^*$ Fixed\\
\end{table*}

\begin{figure}

\epsfig{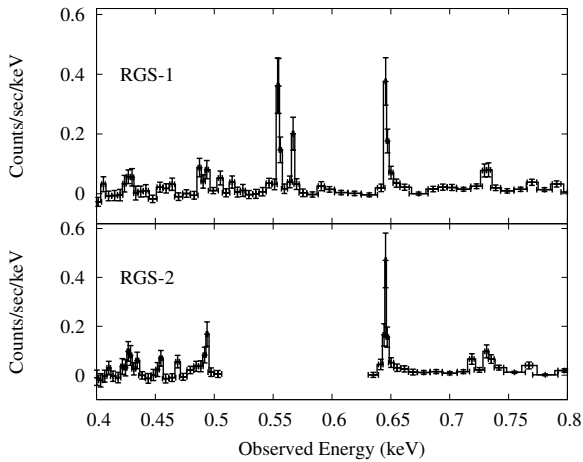}

\caption{\label{rgs}The RGS spectrum of Mrk~3 between 0.4 and 0.8 keV.}

\end{figure}

The analysis of the RGS spectra shows that the soft X-ray EPIC pn spectrum of Mrk~3 is dominated by emission lines, in particular from {O\,\textsc{vii}} and {O\,\textsc{viii}} K$\alpha$ between 0.5-0.7 keV (see Fig. \ref{rgs}). Therefore, we included these lines in the fit with the pn data and added a further powerlaw component to model the soft X-ray continuum. There is no need now for a very steep powerlaw, as was found in BeppoSAX \citep{cappi99} and in our preliminary analysis of the pn spectrum. Instead, a powerlaw with the same photon index of the primary continuum best fits the continuum underlying both the RGS and the EPIC pn emission line spectra. This component is most likely due to scattering of the primary continuum by a photoionised, Compton-thin gas. Indeed, an estimate of the column density of this material can be calculated on the basis of the ratio between the fluxes of the reflection component and the primary emission, which is $\simeq1\%$. Assuming a covering factor of 0.5, just to have an order of magnitude estimate, the column density of the photoionised gas would be approximately $3\times10^{22}$ cm$^{-2}$. As already noted by \citet{imf02} and \citet[][but see also references intherein]{gua04}, it may well be that low resolution spectra of highly absorbed Seyfert galaxies are often affected by a blending of strong emission lines which mimic a continuum component, and only high resolution spectroscopy can unveil the real nature of this kind of soft excess.

Table \ref{lines} lists all the emission lines likely produced in this material which were included in the fit, both in the RGS and in the EPIC pn spectra. All the line widths are unresolved, so they are assumed as $\delta$ functions. Along with lines identified with H-like emission from N, O,  Ne, Mg and Si, lines compatible with blends of He-like complexes of the same ionic species are detected. However, the EPIC pn moderate energy resolution does not allow to resolve any of them, while only the {O\,\textsc{vii}}, {Ne\,\textsc{ix}} and {Mg\,\textsc{xi}} systems are actually separated into their forbidden and resonant components in the RGS spectra. In these cases (taking tentatively also into account the EPIC pn centroid energies of the blends), the results suggest an important role of the forbidden line, which is usually interpreted as the signature of emission predominantly from photoionised rather than collisional plasmas \citep[e.g. ][]{pd00}. Moreover, the very presence of the {O\,\textsc{vii}} and {O\,\textsc{viii}} Radiative Recombination Continua (RRC) argues against a collisional plasma, since in that case this feature would be much broader and very difficult to detect \citep{lied99,lp96}.

In the \textit{Chandra} HETG data, some of the He--like complexes are resolved, showing possibly a larger contribution from the resonant line (except for {O\,\textsc{vii}}) that was interpreted in terms of photoexcitation of resonant transitions, expected to occur in photoionised plasmas \citep{sako00b}. These authors argued against the presence of an additional collisionally ionized plasma component (that could explain the anomalous strong resonance lines) because strong Fe L-shell emission is lacking, a result which appears to be confirmed by the XMM--Newton spectrum.

Therefore, plasma diagnostics with both \textit{Chandra} and XMM--\textit{Newton} seems to confirm that emission from a photoionised material dominates the soft X--ray spectrum of Mrk~3. The (tentative) difference between the role of forbidden and resonance lines in the two spectra may be due to the spatial resolution of the two instruments, which could also play a relevant role, given that the \textit{Chandra} observation showed extended emission which is not spatially resolved by XMM-\textit{Newton}. The XMM-\textit{Newton} data might simply indicate the importance of photoionisation not only in the innermost nuclear region (as shown by \textit{Chandra} data) but also at larger radii, corresponding to the extended emission revealed by \textit{Chandra}. In the nucleus, low column densities may be the origin of stronger resonant lines, while larger column densities in the extended region could be responsible for predominant forbidden lines \citep[see e.g.][for a full discussion on the relative role of the He-like transitions as a function of the column density]{bianchi05}.

\subsubsection{The ionised iron lines}

\begin{table}
\caption{\label{hotlines}List of the emission lines included in the best fit model, originating from a highly ionised material.}
\begin{center}
\begin{tabular}{cccc}
\textbf{Energy (keV)} & \textbf{Flux$^a$} & \textbf{Id.} & $\mathbf{E_T}$ $^b$ \textbf{(keV)}\\
\hline
& & & \\
\multirow{3}*{$6.71^{+0.03}_{-0.02}$} & \multirow{3}*{$0.4\pm0.2$} &\multirow{3}*{{Fe\,\textsc{xxv}} K$\alpha$} & 6.637 (f)\\
& & & 6.675 (i)\\
& & & 6.700 (r)\\
& & &\\
$6.97^*$ & $<0.1$ & {Fe\,\textsc{xxvi}} K$\alpha$ & 6.966\\
& & & \\
$7.60\pm0.05$ & $0.4^{+0.1}_{-0.2}$ & {Ni\,\textsc{xxi-xxiv}} K$\alpha$ & 7.550-7.668\\
& & &\\
\hline
\end{tabular}
\end{center}
$^a$ $10^{-5}$ ph cm$^{-2}$ s$^{-1}$

$^b$ Theoretical energies for the transitions. For He-like ions, forbidden, intercombination and resonant lines are separately listed \citep[data from NIST: ][ and references intherein]{nist}

$^*$ Fixed
\end{table}

A {Fe\,\textsc{xxv}} K$\alpha$ line is required in the EPIC pn spectrum, with a centroid energy of $6.71^{+0.03}_{-0.02}$ keV (see Table \ref{hotlines}). This line is actually composed of four transitions, the resonance line (\textit{w}: 6.700 keV), two intercombination lines (\textit{x} and \textit{y}: mean energy 6.675 keV) and the forbidden line (\textit{z}: 6.637 keV). The best fit energy in our data suggests that the dominant transition is the resonance \textit{w}. This is generally taken as a sign that the gas is mainly in collisional equilibrium \citep[see e.g.][]{pd00}. However, the \textit{w} line can be significantly enhanced by resonant scattering, which is expected to occur in photoionised plasma. This process is very effective at low column densities and the resulting resonance line becomes the strongest in the He-like iron spectrum \citep[see e.g.][and references therein]{mbf96,bianchi05}. In NGC~1068, the {Fe\,\textsc{xxv}} \textit{w} and \textit{z} lines were actually resolved and the large ratio between the former to the latter allowed \citet{bianchi05} to estimate a column density of a few $10^{21}$ cm$^{-2}$. A similar estimate can be done for Mrk~3, if the centroid energy of the blend of the He-like iron lines is indeed interpreted as the result of a dominant \textit{w} line.

A line from highly ionised Ni is also detected, possibly associated with the same reflector, while we only found an upper limit for a {Fe\,\textsc{xxvi}} line (see Table \ref{lines}). A much stronger He-like iron line suggests that the ionisation parameter of the gas is likely lower than $\log U_x\simeq-0.5$ \citep[see][ for details]{bianchi05}. On the other hand, the ionisation parameter appropriate for the observed {Fe\,\textsc{xxv}} is much higher than the one consistent with the production of the other lines found in the soft X-ray spectrum, requiring at least two different ionised reflecting materials in the circumnuclear region of Mrk~3, similarly to what found, for example, for NGC~1068 \citep[][and references therein]{matt04}.

\section{IXO~30}
\label{ixo30}

\begin{figure*}

\epsfig{file=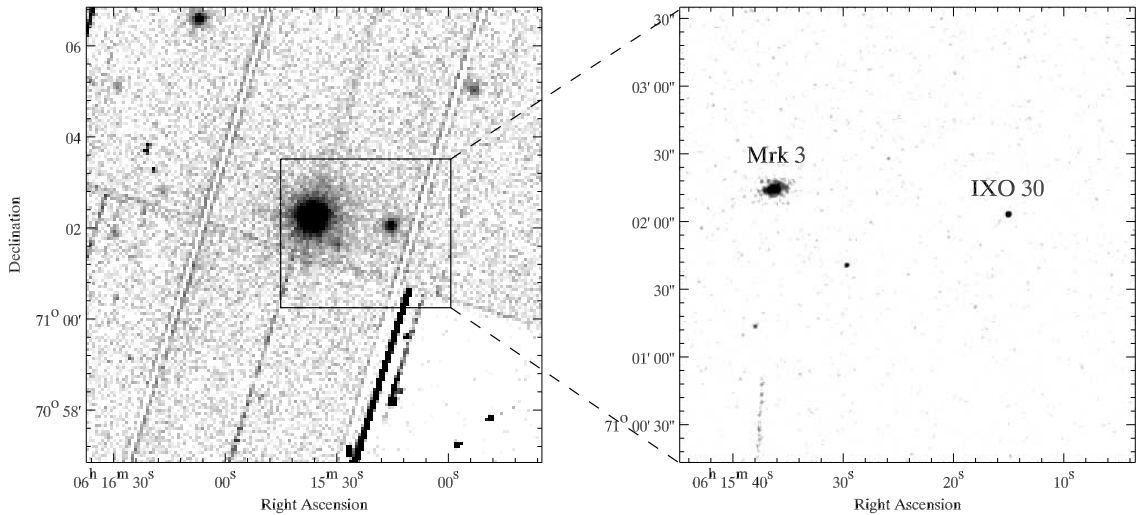, width=15cm}

\caption{\label{ixo30image}\textit{Left}: EPIC pn image of the Mrk~3 field. \textit{Right}: \textit{Chandra} image of the rectangular region shown aside, together with the identification of Mrk~3 and IXO~30.}

\end{figure*}

As shown in figure \ref{ixo30image}, a bright source is apparent $\simeq1.6$ arcmin southeast of the nucleus. The much better astrometry of \textit{Chandra} allows us to locate it at $\alpha_{2000}=06^h15^m15^s$, $\delta_{2000}=+71\degr 02\arcmin 05\arcsec$, fully consistent with \textit{ROSAT} source IXO~30 \citep{cp02}.

\begin{figure}

\epsfig{file=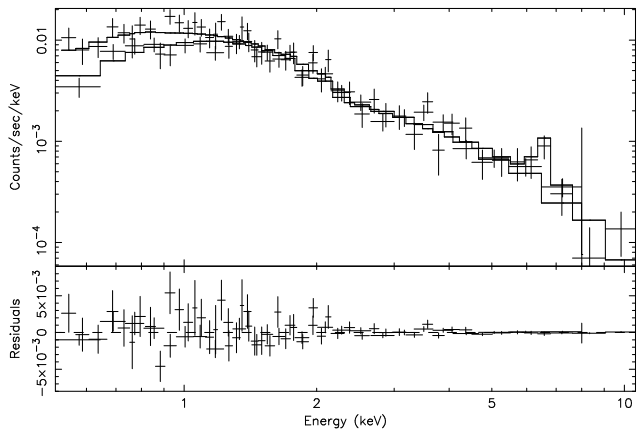, width=5.6cm, angle=-90}

\caption{\label{ixo30fit}The EPIC pn and MOS1+MOS2 spectrum, best fit model and residuals for IXO~30.}

\end{figure}

This source was first detected by \textit{ROSAT} and discussed by \citet{tum93} and \citet{morse95}.
We present here the first X-ray spectrum for IXO~30. A combined pn-MOS fit with a simple absorbed powerlaw leads to an acceptable fit ($\chi^2=86/75$ d.o.f.), but the addition of an unresolved ($\sigma<600$ eV) gaussian line at $6.5^{+0.3}_{-0.2}$ keV is required at the 97\% confidence level, according to F-test. The flux of this feature, likely identified as an iron line, even if the large errors prevent us from stating whether it is ionised or not, is $(8\pm6)\times10^{-7}$ ph cm$^{-2}$ s$^{-1}$, corresponding to a very large EW of $1.0\pm0.8$ keV. The final fit is good ($\chi^2=76/72$ d.o.f.), with $N_{H}=(1.7\pm0.4)\times10^{21}$ cm$^{-2}$ and $\Gamma=1.77^{+0.13}_{-0.10}$ (see figure \ref{ixo30fit}). A fit of comparable quality ($\chi^2=77/73$ d.o.f.) is achieved by a bremsstrahlung component at a temperature of $7.5^{+2.1}_{-1.6}$ keV, absorbed only by the Galactic column density, plus an iron line with the same properties as for the previous model. Finally, we have also tried a single temperature plasma component (model \textsc{mekal}, assuming Solar metallicity), which gives a good fit ($\chi^2=83/76$ d.o.f.) with a temperature, $8.2^{+1.9}_{-1.2}$ keV, which is enough to take into account the iron line without any further component.

\begin{figure}
\begin{center}
\epsfig{file=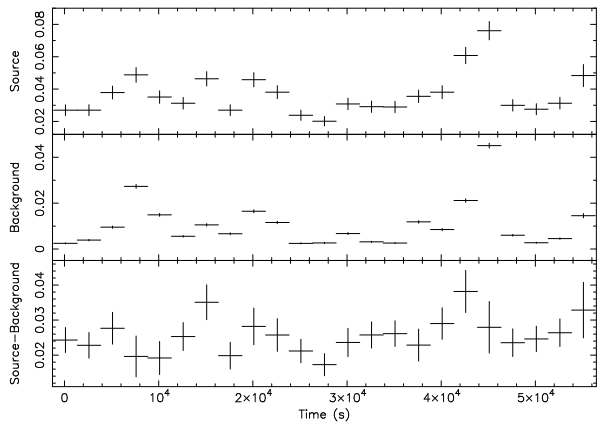, width=5.5cm, angle=-90}
\end{center}

\caption{\label{ixo30_lc}\textit{Upper}: Lightcurve of the adopted extraction region for IXO~30. \textit{Middle}: Lightcurve for the background, with count rates rescaled to the area of the source extraction region. \textit{Lower}: Background-subtracted lightcurve of IXO~30.}

\end{figure}

\begin{figure}
\begin{center}
\epsfig{file=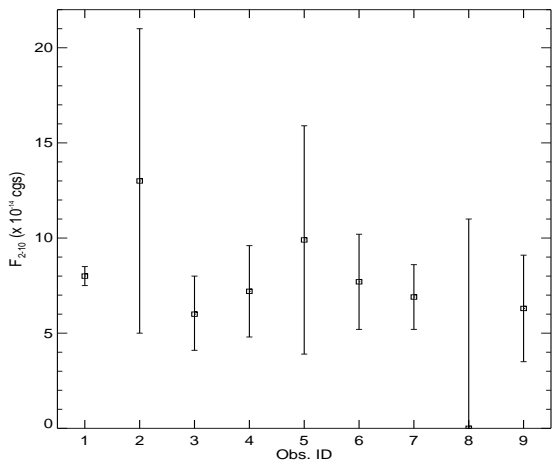, width=7.2cm, height=6cm}
\end{center}

\caption{\label{ixo30_flux}The history of the observed flux of IXO~30 among the available EPIC pn observations (listed in Table \ref{ixo30_obs}). All spectra were fitted with an absorbed powerlaw, adopting the best fit values of the longest observation.}

\end{figure}

The observed 2-10 keV flux (for all the models) is $\simeq8.5\times10^{-14}$ erg cm$^{-2}$ s$^{-1}$, corresponding to a luminosity of around $3.5\times10^{40}$ erg s$^{-1}$, if IXO~30 lies at the same distance of Mrk~3, in good agreement with what reported by \citet{tum93} for the \textit{ROSAT} observation. We found a somewhat higher flux, $\simeq1.2\times10^{-13}$ erg s$^{-1}$, in the \textit{Chandra} spectrum of the source, but the fit parameters are loosely constrained by the poor statistics, being $N_{H}=(1.9^{+1.8}_{-1.1})\times10^{21}$ cm$^{-2}$ and $\Gamma=1.9^{+0.4}_{-0.3}$ for the powerlaw model, with an upper limit of 1.3 keV to the EW of the iron line. On the other hand, the analysis of all the available EPIC pn observations does not display any significant variability of the source flux within a time span of almost two years (see Fig. \ref{ixo30_flux} and Table \ref{ixo30_obs}). However, it is interesting to note that the only observation long enough to allow some spectral analysis (observation 3 in Table \ref{ixo30_obs}) suggests a possible variation of the absorbing column density, since a larger value, $N_\mathrm{H}=7.2^{+4.9}_{-3.6}\times10^{21}$ cm$^{-2}$, is required by the data. Finally, no clear evidence for short-term varibility is present in the EPIC pn lightcurve of IXO~30 (see figure \ref{ixo30_lc}), considering that the high background periods (filtered for the extraction of the spectra) are comparable to the source flux and are not easily subtracted from the lightcurve.

\begin{table}

\caption{\label{ixo30_obs}Log of all the EPIC pn observations of IXO~30 analysed in this paper. See text for details.}

\begin{center}
\begin{tabular}{|c|c|c|c|}
\hline
\textbf{Obs. ID} & \textbf{Date} & \textbf{Exp. Time (ks)}\\
1 & 2000-10-19 & 53\\
2 & 2001-03-12 & 2\\
3 & 2001-03-20 & 10\\
4 & 2001-03-28 & 5\\
5 & 2001-04-05 & 3\\
6 & 2002-03-10 & 5\\
7 & 2002-03-25 & 5\\
8 & 2002-04-18 & 1\\
9 & 2002-09-16 & 5\\
\hline
\end{tabular}
\end{center}

\end{table}

We have searched for an optical counterpart of IXO~30 in the Digitized Sky Survey, but we did not find any possible companion neither in the POSS-II Red nor in the Blue plate, so implying that the source is fainter than R~$\simeq21$ and Bj~$\simeq22.5$. It is therefore quite difficult to guess its real nature. A possible interpretation is in terms of an AGN, which must be a background object, otherwise its luminosity would be too low. In any case, the presence of an iron line allows us to put an upper limit on the redshift of the source, assuming that the line comes from H-like iron: this limit, being approximately $z<0.1$, implies that the 2-10 keV luminosity of the source cannot be larger than $\simeq2\times10^{42}$ erg s$^{-1}$, quite low for an AGN. Moreover, the EW of the iron line, though determined with large uncertainty, would be typical of a Compton-thick object, but this would be at odds with the observed photon index.

On the other hand, if associated with the Mrk~3 galaxy, its luminosity is comparable with that usually found for Ultra Luminous X-ray sources (ULXs) often detected in the vicinity of an AGN and possibly related to Intermediate-Mass Black Holes \citep[see e.g.][]{cm04}. If the latter interpretation is correct, at least a mass of $\simeq300$ M$_\odot$ is required, if the source is emitting near the Eddington limit. However, the origin of the huge iron line remains obscure, even if at least another ULX was recently found in M82, showing a similar feature \citep{sm03}. Furthermore, a separation of 1.6 arcmin at the redshift of Mrk~3 would imply a large distance of the source from the nucleus, being $\simeq25$ kpc, which is very close to the optical $D_{25}$ diameter of the host galaxy \citep{devau91}, so that the probability of contamination from background/foreground objects is significant \citep[see e.g.][]{pc04}.

Finally, it is possible an interpretation in terms of a Galactic source, like a Cataclysmic Variable (CV). This scenario is favoured by the spectral analysis, since a bremsstrahlung component and a huge iron line are generally the main ingredients to model these objects \citep[see e.g.][]{ms93}. At a distance of $\simeq500$ pc from the Earth, a 2-10 keV luminosity of $\simeq2\times10^{30}$ erg s$^{-1}$ would make IXO~30 a weak CV, but still in the observed range \citep[see e.g.][]{ms93}. On the other hand, if we assume, as it is generally found, that the secondary of this system is a typical M red dwarf, we can get a lower limit on the distance of the CV, based on the upper limit we have on its apparent R magnitude. The resulting distance, approximately $d>4$ kpc, is in agreement with an origin of the source in the Galactic disc, taking also into account its Galactic coordinates. This would imply a 2-10 keV luminosity larger than $10^{32}$ erg s$^{-1}$, which is within the observed range in CVs.

\section{Conclusions}

We have analysed the first XMM-\textit{Newton} spectrum of Mrk~3. The analysis confirm previous results, showing a spectrum composed by three main components: a strongly absorbed powerlaw, a pure Compton reflection with the same photon index associated to an iron K$\alpha$ line, both produced as reflection from a Compton-thick torus, and an unabsorbed powerlaw, again with the same photon index of the primary continuum, associated to a large number of emission lines, likely produced as reflection from a Compton-thin, photoionised material.

The iron K$\alpha$ line EW with respect to the Compton reflection component, being only $610^{+30}_{-50}$ eV, is consistent with a low inclination angle and an iron underabundance of a factor $\simeq0.82$, properties indipendently derived for the torus, via the amount of Compton reflection and the depth of the iron edge. The iron line width is actually resolved in the EPIC pn spectrum, with $\sigma=32^{+13}_{-14}$ eV, corresponding to FWHM=$3\,500^{+1\,400}_{-1\,500}$ km s$^{-1}$ if produced by Doppler broadening, thus putting an estimate to the inner radius of the torus, $r=0.6^{+1.3}_{-0.3}\,\sin^2{i}$ pc. Moreover, a possible variation on a timescale of $\simeq7$ months of the ionisation stage of iron between the \textit{Chandra} and the XMM-\textit{Newton} observations is fully compatible with the photoionisation time for iron at a distance of around 1 pc.

The soft X-ray spectrum of Mrk~3 is dominated by emission lines of H-- and He--like from the most abundant metals, superimposed over an unabsorbed powerlaw with the same photon index of the primary continuum. It is important to note that the full resolution of these lines in the RGS spectra was required before correctly fitting this part of the spectrum in the lower resolution pn data, thus preventing the adoption of a much steeper ($\Gamma\simeq3$) powerlaw, whose physical interpretation would have been less straightforward. From the ratio between the fluxes of the reflected component and the primary continuum, a column density of a few $10^{22}$ cm$^{-2}$ can be derived for the photoionised material. However, it must be noted that at least two different ionised reflectors are needed to take into account the emission lines from lighter metals and the {Fe\,\textsc{xxv}} emission line at $6.71^{+0.03}_{-0.02}$ keV.

We have also presented the first X-ray spectrum of \textit{ROSAT} source IXO~30, which shows a huge iron line at $6.5^{+0.3}_{-0.2}$ keV and is well represented either by an absorbed powerlaw with $\Gamma\simeq1.8$ or bremsstrahlung emission at a temperature of $7.5^{+2.1}_{-1.6}$ keV. Even if the lack of any optical counterpart precludes excluding other possibilities, such as an ULX at the distance of Mrk~3, its spectral properties point to a likely identification in terms of a Galactic CV.

\section*{Acknowledgements}

We would like to thank G. Matt and M. Guainazzi for many useful discussions, and the referee for his valuable comments. SB acknowledges Fondazione Della Riccia for some financial support. GM and KI thank the PPARC for support. ACF thanks the Royal Society for support. This paper is based on observations obtained with XMM-\textit{Newton}, an ESA science mission with instruments and contributions directly funded by ESA Member States and the USA (NASA). The Digitized Sky Survey was produced at the Space Telescope Science Institute under U.S. Government grant NAG W-2166. The images of these surveys are based on photographic data obtained using the Oschin Schmidt Telescope on Palomar Mountain and the UK Schmidt Telescope.

\bibliographystyle{mn}
\bibliography{sbs}

\begin{thebibliography}{45}
\expandafter\ifx\csname natexlab\endcsname\relax\def\natexlab#1{#1}\fi

\bibitem[{{Anders} \& {Grevesse}(1989)}]{ag89}
{Anders} E., {Grevesse} N., 1989, \gca, 53, 197

\bibitem[{{Awaki} {et~al.}(1991){Awaki}, {Koyama}, {Inoue}, \&
  {Halpern}}]{awaki91}
{Awaki} H., {Koyama} K., {Inoue} H., {Halpern} J.~P., 1991, \pasj, 43, 195

\bibitem[{{Awaki} {et~al.}(1990){Awaki}, {Koyama}, {Kunieda}, \&
  {Tawara}}]{awaki90}
{Awaki} H., {Koyama} K., {Kunieda} H., {Tawara} Y., 1990, \nat, 346, 544

\bibitem[{{Basko}(1978)}]{basko78}
{Basko} M.~M., 1978, \apj, 223, 268

\bibitem[{{Bearden}(1967)}]{bea67}
{Bearden} J.~A., 1967, Reviews of Modern Physics, 39, 78

\bibitem[{{Bianchi} {et~al.}(2002){Bianchi}, {Matt}, {Fiore}, {Fabian},
  {Iwasawa}, \& {Nicastro}}]{bianchi02}
{Bianchi} S., {Matt} G., {Fiore} F., {Fabian} A.~C., {Iwasawa} K., {Nicastro}
  F., 2002, \aap, 396, 793

\bibitem[{{Bianchi} {et~al.}(2005){Bianchi}, {Matt}, {Nicastro}, {Porquet}, \&
  {Dubau}}]{bianchi05}
{Bianchi} S., {Matt} G., {Nicastro} F., {Porquet} D., {Dubau} J., 2005, \mnras,
  357, 599

\bibitem[{{Cappi} {et~al.}(1999){Cappi}, {Bassani}, {Comastri}, {Guainazzi},
  {Maccacaro}, {Malaguti}, {Matt}, {Palumbo}, {Blanco}, {Dadina}, {dal Fiume},
  {di Cocco}, {Fabian}, {Frontera}, {Maiolino}, {Piro}, {Trifoglio}, \&
  {Zhang}}]{cappi99}
{Cappi} M. et al., 1999, \aap, 344, 857

\bibitem[{{Colbert} \& {Miller}(2004)}]{cm04}
{Colbert} E.~J.~M., {Miller} M.~C., 2004, Invited review talk at the Tenth
  Marcel Grossmann Meeting on General Relativity, Rio de Janeiro, July 20-26,
  2003. Proceedings edited by M. Novello, S. Perez-Bergliaffa and R. Ruffini,
  World Scientific, Singapore, 2005 (astro-ph/0402677)

\bibitem[{{Colbert} \& {Ptak}(2002)}]{cp02}
{Colbert} E.~J.~M., {Ptak} A.~F., 2002, \apjs, 143, 25

\bibitem[{{de Vaucouleurs} {et~al.}(1991){de Vaucouleurs}, {de Vaucouleurs},
  {Corwin}, {Buta}, {Paturel}, \& {Fouque}}]{devau91}
{de Vaucouleurs} G., {de Vaucouleurs} A., {Corwin} H.~G., {Buta} R.~J.,
  {Paturel} G., {Fouque} P., 1991, {Third Reference Catalogue of Bright
  Galaxies}. Volume 1-3, XII, 2069 pp.~7 figs..~ Springer-Verlag Berlin
  Heidelberg New York

\bibitem[{{Dickey} \& {Lockman}(1990)}]{dl90}
{Dickey} J.~M., {Lockman} F.~J., 1990, \araa, 28, 215

\bibitem[{{Georgantopoulos} {et~al.}(1999){Georgantopoulos}, {Papadakis},
  {Warwick}, {Smith}, {Stewart}, \& {Griffiths}}]{georg99}
{Georgantopoulos} I., {Papadakis} I., {Warwick} R.~S., {Smith} D.~A., {Stewart}
  G.~C., {Griffiths} R.~G., 1999, \mnras, 307, 815

\bibitem[{{Guainazzi} {et~al.}(2004){Guainazzi}, {Rodriguez-Pascual}, {Fabian},
  {Iwasawa}, \& {Matt}}]{gua04}
{Guainazzi} M., {Rodriguez-Pascual} P., {Fabian} A.~C., {Iwasawa} K., {Matt}
  G., 2004, \mnras, 355, 297

\bibitem[{{House}(1969)}]{house69}
{House} L.~L., 1969, \apjs, 18, 21

\bibitem[{{Iwasawa} {et~al.}(2002){Iwasawa}, {Maloney}, \& {Fabian}}]{imf02}
{Iwasawa} K., {Maloney} P.~R., {Fabian} A.~C., 2002, \mnras, 336, L71

\bibitem[{{Iwasawa} {et~al.}(1994){Iwasawa}, {Yaqoob}, {Awaki}, \&
  {Ogasaka}}]{iwa94}
{Iwasawa} K., {Yaqoob} T., {Awaki} H., {Ogasaka} Y., 1994, \pasj, 46, L167

\bibitem[{{Kaastra} \& {Mewe}(1993)}]{km93}
{Kaastra} J.~S., {Mewe} R., 1993, \aaps, 97, 443

\bibitem[{{Kruper} {et~al.}(1990){Kruper}, {Canizares}, \& {Urry}}]{kruper90}
{Kruper} J.~S., {Canizares} C.~R., {Urry} C.~M., 1990, \apjs, 74, 347

\bibitem[{{Liedahl}(1999)}]{lied99}
{Liedahl} D.~A., 1999, Lecture Notes in Physics, Berlin Springer Verlag, 520,
  189

\bibitem[{{Liedahl} \& {Paerels}(1996)}]{lp96}
{Liedahl} D.~A., {Paerels} F., 1996, \apjl, 468, L33+

\bibitem[{{Magdziarz} \& {Zdziarski}(1995)}]{mz95}
{Magdziarz} P., {Zdziarski} A.~A., 1995, \mnras, 273, 837

\bibitem[{{Matt}(2002)}]{matt02}
{Matt} G., 2002, \mnras, 337, 147

\bibitem[{{Matt} {et~al.}(2004){Matt}, {Bianchi}, {Guainazzi}, \&
  {Molendi}}]{matt04}
{Matt} G., {Bianchi} S., {Guainazzi} M., {Molendi} S., 2004, \aap, 414, 155

\bibitem[{{Matt} {et~al.}(1996){Matt}, {Brandt}, \& {Fabian}}]{mbf96}
{Matt} G., {Brandt} W.~N., {Fabian} A.~C., 1996, \mnras, 280, 823

\bibitem[{{Miller} \& {Goodrich}(1990)}]{mg90}
{Miller} J.~S., {Goodrich} R.~W., 1990, \apj, 355, 456

\bibitem[{{Molendi} {et~al.}(2003){Molendi}, {Bianchi}, \& {Matt}}]{mbm03}
{Molendi} S., {Bianchi} S., {Matt} G., 2003, \mnras, 343, L1

\bibitem[{{Morse} {et~al.}(1995){Morse}, {Wilson}, {Elvis}, \&
  {Weaver}}]{morse95}
{Morse} J.~A., {Wilson} A.~S., {Elvis} M., {Weaver} K.~A., 1995, \apj, 439, 121

\bibitem[{{Mukai} \& {Shiokawa}(1993)}]{ms93}
{Mukai} K., {Shiokawa} K., 1993, \apj, 418, 863

\bibitem[{{Ogle} {et~al.}(2003){Ogle}, {Brookings}, {Canizares}, {Lee}, \&
  {Marshall}}]{ogle03}
{Ogle} P.~M., {Brookings} T., {Canizares} C.~R., {Lee} J.~C., {Marshall} H.~L.,
  2003, \aap, 402, 849

\bibitem[{{Palmeri} {et~al.}(2003){Palmeri}, {Mendoza}, {Kallman}, {Bautista},
  \& {Mel{\' e}ndez}}]{palm03}
{Palmeri} P., {Mendoza} C., {Kallman} T.~R., {Bautista} M.~A., {Mel{\' e}ndez}
  M., 2003, \aap, 410, 359

\bibitem[{{Piconcelli} {et~al.}(2004){Piconcelli}, {Jimenez-Bail{\' o}n},
  {Guainazzi}, {Schartel}, {Rodr{\'{\i}}guez-Pascual}, \& {Santos-Lle{\'
  o}}}]{pico04}
{Piconcelli} E., {Jimenez-Bail{\' o}n} E., {Guainazzi} M., {Schartel} N.,
  {Rodr{\'{\i}}guez-Pascual} P.~M., {Santos-Lle{\' o}} M., 2004, \mnras, 351,
  161

\bibitem[{{Porquet} \& {Dubau}(2000)}]{pd00}
{Porquet} D., {Dubau} J., 2000, \aaps, 143, 495

\bibitem[{{Ptak} \& {Colbert}(2004)}]{pc04}
{Ptak} A., {Colbert} E., 2004, \apj, 606, 291

\bibitem[{{Ralchenko} {et~al.}(2004){Ralchenko}, {Jou}, {Kelleher}, {Kramida},
  {Musgrove}, {Reader}, {Wiese}, \& {Olsen}}]{nist}
{Ralchenko} Y., {Jou} F.-C., {Kelleher} D.~E., {Kramida} A.~E., {Musgrove} A.,
  {Reader} J., {Wiese} W., {Olsen} K., 2004, {NIST Atomic Spectra Database
  (version 3.0-beta)}, Available: http://physics.nist.gov/asd3. National
  Institute of Standards and Technology, Gaithersburg, MD.

\bibitem[{{Reynolds} {et~al.}(1995){Reynolds}, {Fabian}, {Nandra}, {Inoue},
  {Kunieda}, \& {Iwasawa}}]{rey95}
{Reynolds} C.~S., {Fabian} A.~C., {Nandra} K., {Inoue} H., {Kunieda} H.,
  {Iwasawa} K., 1995, \mnras, 277, 901

\bibitem[{{Sako} {et~al.}(2000){Sako}, {Kahn}, {Paerels}, \&
  {Liedahl}}]{sako00b}
{Sako} M., {Kahn} S.~M., {Paerels} F., {Liedahl} D.~A., 2000, \apjl, 543, L115

\bibitem[{{Strohmayer} \& {Mushotzky}(2003)}]{sm03}
{Strohmayer} T.~E., {Mushotzky} R.~F., 2003, \apjl, 586, L61

\bibitem[{{Str{\"u}der} {et~al.}(2001){Str{\"u}der}, {Briel}, {Dennerl},
  {Hartmann}, {Kendziorra}, {Meidinger}, {Pfeffermann}, {Reppin}, {Aschenbach},
  {Bornemann}, {Br{\" a}uninger}, {Burkert}, {Elender}, {Freyberg}, {Haberl},
  {Hartner}, {Heuschmann}, {Hippmann}, {Kastelic}, {Kemmer}, {Kettenring},
  {Kink}, {Krause}, {M{\" u}ller}, {Oppitz}, {Pietsch}, {Popp}, {Predehl},
  {Read}, {Stephan}, {St{\" o}tter}, {Tr{\" u}mper}, {Holl}, {Kemmer},
  {Soltau}, {St{\" o}tter}, {Weber}, {Weichert}, {von Zanthier},
  {Carathanassis}, {Lutz}, {Richter}, {Solc}, {B{\" o}ttcher}, {Kuster},
  {Staubert}, {Abbey}, {Holland}, {Turner}, {Balasini}, {Bignami}, {La
  Palombara}, {Villa}, {Buttler}, {Gianini}, {Lain{\' e}}, {Lumb}, \&
  {Dhez}}]{struder01}
{Str{\"u}der} L. et al., 2001, \aap, 365, L18

\bibitem[{{Tran}(1995)}]{tran95}
{Tran} H.~D., 1995, \apj, 440, 565

\bibitem[{{Turner} {et~al.}(2001){Turner}, {Abbey}, {Arnaud}, {Balasini},
  {Barbera}, {Belsole}, {Bennie}, {Bernard}, {Bignami}, {Boer}, {Briel},
  {Butler}, {Cara}, {Chabaud}, {Cole}, {Collura}, {Conte}, {Cros}, {Denby},
  {Dhez}, {Di Coco}, {Dowson}, {Ferrando}, {Ghizzardi}, {Gianotti}, {Goodall},
  {Gretton}, {Griffiths}, {Hainaut}, {Hochedez}, {Holland}, {Jourdain},
  {Kendziorra}, {Lagostina}, {Laine}, {La Palombara}, {Lortholary}, {Lumb},
  {Marty}, {Molendi}, {Pigot}, {Poindron}, {Pounds}, {Reeves}, {Reppin},
  {Rothenflug}, {Salvetat}, {Sauvageot}, {Schmitt}, {Sembay}, {Short},
  {Spragg}, {Stephen}, {Str{\" u}der}, {Tiengo}, {Trifoglio}, {Tr{\" u}mper},
  {Vercellone}, {Vigroux}, {Villa}, {Ward}, {Whitehead}, \& {Zonca}}]{turner01}
{Turner} M.~J.~L. et al., 2001, \aap, 365, L27

\bibitem[{{Turner} {et~al.}(1993){Turner}, {Urry}, \& {Mushotzky}}]{tum93}
{Turner} T.~J., {Urry} C.~M., {Mushotzky} R.~F., 1993, \apj, 418, 653

\bibitem[{{Verner} \& {Yakovlev}(1995)}]{vy95}
{Verner} D.~A., {Yakovlev} D.~G., 1995, \aaps, 109, 125

\bibitem[{{Woo} \& {Urry}(2002)}]{wu02}
{Woo} J., {Urry} C.~M., 2002, \apj, 579, 530

\bibitem[{{Yaqoob} {et~al.}(2001){Yaqoob}, {George}, {Nandra}, {Turner},
  {Serlemitsos}, \& {Mushotzky}}]{yaq01}
{Yaqoob} T., {George} I.~M., {Nandra} K., {Turner} T.~J., {Serlemitsos} P.~J.,
  {Mushotzky} R.~F., 2001, \apj, 546, 759

\end{thebibliography}

\label{lastpage}

\end{document}